\documentclass[12pt,a4paper,nofootinbib]{revtex4}
\usepackage{multirow}
\usepackage{amssymb}
\usepackage{amsmath,graphicx}
\usepackage{array}
\usepackage{braket}
\usepackage{bm}
\usepackage{enumerate}
\usepackage{epstopdf}
\usepackage{amsmath}
\usepackage{dcolumn,caption,booktabs}
\usepackage{slashed}
\usepackage{amsfonts}
\usepackage{diagbox}
\usepackage{adjustbox}
\usepackage{tcolorbox}
\usepackage{float}
\usepackage{collectbox}
\usepackage{listings}
\usepackage[unicode]{hyperref}
\usepackage[mathscr]{eucal}
\usepackage{graphics}
\usepackage{subcaption}
\newcommand{\hs}{\hspace*{0.3cm}}

\newcommand{\be}{\begin{equation}}
	\newcommand{\ee}{\end{equation}}
\newcommand{\bea}{\begin{eqnarray}}
	\newcommand{\eea}{\end{eqnarray}}
\newcommand{\ben}{\begin{enumerate}}
	\newcommand{\een}{\end{enumerate}}
\newcommand{\bde}{\begin{widetext}}
	\newcommand{\ede}{\end{widetext}}

\newcommand{\crn}{\nonumber \\}

\newcommand{\al}{\alpha}
\newcommand{\la}{\lambda}

\newcommand{\va}{\varphi}

\newcommand{\fr}{\frac}

\newcommand{\bc}{\begin{center}}
	\newcommand{\ec}{\end{center}}

\newcommand{\ep}{\epsilon}

\newcommand{\ka}{\kappa}
\newcommand{\La}{\Lambda}
\newcommand{\si}{\sigma}

\begin{document}

\title{Constraint on the Higgs-Dilaton potential via Warm inflation in Two-Time Physics}
\author{Vo Quoc Phong$^{a,b}$}
\email{vqphong@hcmus.edu.vn}
\affiliation{$^a$ Department of Theoretical Physics, University of Science, Ho Chi Minh City 700000, Vietnam\\
	$^b$ Vietnam National University, Ho Chi Minh City 700000, Vietnam}
\author{Ngo Phuc Duc Loc$^{a,b}$}
\email{locngo148@gmail.com}
\affiliation{$^a$ Department of Theoretical Physics, University of Science, Ho Chi Minh City 700000, Vietnam\\
	$^b$ Vietnam National University, Ho Chi Minh City 700000, Vietnam}	

\begin{abstract}
Within the $SP(2,R)$ symmetry, the Two-time model (2T model) has six dimensions with two dimensions of time and the dilaton field that can be identified as inflaton in a warm inflation scenario with potential of the form $\sim\phi^4$. From that consideration, we derive the range of parameters for the Higgs-Dilaton potential, the coupling constant between Higgs and Dialton ($\alpha$) is larger than $0.0053$ and the mass of Dilaton is smaller than $10^{-7}$ GeV. Therefore, the 2T-model indirectly suggests that extra-dimension can  also  be a source of inflation.
\end{abstract}

\maketitle
\section{Introduction}\label{secInt}

From quantum mechanics, there is a symmetry between coordinates and momentums that can be described by a symplectic group. When we combine this spatial symmetry and Poincaré symmetry, it can be generally described by the $Sp(2,R)$ group or leading to the space-time must be larger than the ordinary one, there is exactly an extra dimension of space and time \cite{bars1999,bars2000a}. From there, Two-time physics (2T Physics) was formed. The 2T physics provides a better understanding of symmetries as well as the concept of time.

The 2T model \cite{bars1999,bars2000a,bars2001,survey,bars2006,kuo2006} could be a well known choice. The model presents us with an unusual vision of time that advances on a plane, i.e., it has two dimensions of time. The 2T SM (The Standard Model in 2T Physics) has better features than regular SM in 1T physics. The strong CP problem of QCD has been solved without axion. It shows us the 1T SM as the "shadow" of 2T framework \cite{bars2006,kuo2006}. The concepts of force, particles and their interaction in 2T SM are reduced to 1T Physics through space-time contractions which are examples that clearly show 1T SM (The Standard Model in 1T Physics) fully embedded in 2T SM \cite{bars2006,kuo2006}. The 2T concept can be extended in gravity \cite{survey}. This theory has proposed expanding the number of spacetime dimensions to represent the whole cosmos, as well as the introduction of dilaton and its properties, for more than a decade (since 2008). The remarkable new quantity in the 2T SM model is the Higgs-Dilaton potential which gives us a new type of symmetry breaking \cite{bars2006,kuo2006}. Importantly, the Dilaton potential in this model is open, so it provides us with new bases for solving other dilemmas. In addition, the development directions as well as the success of 2T Physcis can be detailed in Ref.~\cite{bars2001}.

After Alan Guth introduced the concept of inflation in 1981 \cite{8ag}, the nature of the inflaton field responsible for inflation remained elusive. For obvious economical reasons, researchers usually consider Higgs boson as a promising candidate \cite{higgsinf1,higgsinf2,higgsinf3}. Currently, there are many inflation scenarios. The basic scenario is the slow-roll approximation in which the potential changes slowly with respect to the field. We therefore examine this basic approximation in the 2T model through the Higgs-dilaton coupling.

We recognize that the inflation problem has a deep connection with the number of dimensions of space-time. The 2T model therefore has both an extra-dimensional effect and exotic particles but the inflation is not yet analyzed in detail, i.e, extra dimensions and dilaton boson can be new materials for inflation. Therefore, the 2T SM or specifically the Dilaton potential in the 2T SM is chosen to analyze the inflation problem or determine the parameters of this model when considering Dialton as an Inflaton.

This article is organized as follows. In section \ref{sec2}, we give a brief introduction to the 2T model and introduce to a new Dilaton potential. In section \ref{sec3}, we introduce some salient features of warm inflation and show that the dilaton can be identified as inflaton in the warm inflation scenario and get the parameter domain of the Higgs-Dilaton potential. Finally, we summarize and make outlooks in section \ref{sec4}.
\section{A brief introduction to 2T model and a new Dilaton potential}\label{sec2}

\subsection{The $Sp(2,R)$ symmetry in the 2T model}

The main new feature of 2T-physics is a new gauge symmetry called $Sp(2,R)$ that works in phase space $(X^M,P^M)$ (not just in spacetime like 1T-physics). We will show that Sp(2,R gauge symmetry) necessarily requires 2 timelike dimensions.

In 2T-physics, spacetime coordinates and energy-momentum are united in a doublet $(X_1^M,X_2^M)$. $X_1^M=X^M$ is spacetime coordinates in 2T-physics and $X_2^M=P^M$ is energy-momentum in 2T-physics. In short, we can label them as $X_i^M$ with i=1,2. The symplectic transformation will act on this doublet and therefore, spacetime and energy-momentum can transform to each other. This is a brand new property of 2T-physics. In 1T-physics, we just have transformations between spacetime coordinates, for example, the Lorentz transformation or even in gauge theory, the local gauge transformation $e^{i\theta(x^\mu)}$ just involve spacetime coordinates. But in 2T-physics, we treat spacetime coordinates and energy-momentum equally and they can transform to each other.\par
The local symplectic transformation is defined as \cite{18}
\begin{equation}
\delta_\omega X_i^M(\tau)=\epsilon_{ik}\omega^{kl}(\tau)X_l^M(\tau),
\end{equation}
where $\tau$ is the proper time, $\omega^{ij}(\tau)=\omega^{ji}(\tau)$ are parameters of Sp(2,R) and $\epsilon^{ij}$ is tensor Levi-Cavita. To have a connection with the concept of Sp(2,R) group in section 3.1.1, we can write it out explicitly like this
\begin{equation}
\begin{cases}
\delta_\omega X^M(\tau)=\omega^{21}(\tau)X^M(\tau)+\omega^{22}(\tau)P^M(\tau)\\
\delta_\omega P^M(\tau)=-\omega^{11}(\tau)X^M(\tau)-\omega^{12}(\tau)P^M(\tau).
\end{cases}
\end{equation}
Because $X'^M=X^M+\delta_\omega X^M$, $P'^M=P^M+\delta_\omega P^M$ so in the matrix form
\begin{equation}
\begin{pmatrix}
X'^M\\P'^M
\end{pmatrix}
=\begin{pmatrix}
1+\omega^{21}&\omega^{22}\\
-\omega^{11}&1-\omega^{12}
\end{pmatrix}
\begin{pmatrix}
X^M\\P^M
\end{pmatrix}
=S\begin{pmatrix}
X^M\\P^M
\end{pmatrix}.
\end{equation}
So S should be an element of the symplectic group
\begin{align}
\begin{split}
S^TJS&=\begin{pmatrix}
1+\omega^{21}&-\omega^{11}\\
\omega^{22}&1-\omega^{12}
\end{pmatrix}
\begin{pmatrix}
0&1\\
-1&0
\end{pmatrix}
\begin{pmatrix}
1+\omega^{21}&\omega^{22}\\
-\omega^{11}&1-\omega^{12}
\end{pmatrix}\\
&=\begin{pmatrix}
0&1-(\omega^{12})^2+\omega^{11}\omega^{22}\\
-[1-(\omega^{12})^2+\omega^{11}\omega^{22}]&0
\end{pmatrix}.
\end{split}
\end{align}
Therefore, we found the constraints of symplectic parameters $(\omega^{12})^2=\omega^{11}\omega^{22}$ or
\begin{equation}det\begin{pmatrix}
\omega^{11}&\omega^{12}\\
\omega^{21}&\omega^{22}
\end{pmatrix}=0.
\end{equation} 

And remember that $\omega^{12}=\omega^{21}$ by definition. In that case, S is indeed a symplectic element. Meanwhile, the local symplectic transformations for the gauge fields is
\begin{equation}
\delta_\omega A^{ij}=\partial_\tau\omega^{ij}+\omega^{ik}\epsilon_{kl}A^{lj}+\omega^{jk}\epsilon_{kl}A^{il},
\end{equation}
for the gauge fields $A^{ij}=A^{ji}$. 
The covariant derivative is
\begin{equation}
D_\tau X_i^M=\partial_\tau X_i^M-\epsilon_{ik}A^{kl}X_l^M.
\end{equation}
The gauge fields here are not normal gauge fields corresponding to internal transformation of the field. In 1T-physics, in order to make the Lagrangian invariant under local gauge transformations of the scalar field, or spinor,... (we call this type of transformations internal transformations, they transform the fields, not coordinates) we have to replace the normal derivative by the covariant derivative that contains the gauge field and add the kinetic term of the gauge fields (we can not add the mass term since it will not be invariant under the gauge transformation of the gauge field, therefore we need the Higgs mechanism to generate the mass of the gauge fields). In 2T-physics, however, we also have the gauge fields  corresponding to the transformations of the coordinates (we call this type of transformations external transformations). This is a new property that we do not have in 1T-physics.
\par
The Lagrangian that is invariant under this local symplectic transformation is\cite{18}
\begin{align}
\begin{split}
S_0&=\frac{1}{2}\int d\tau (D_\tau X_i^M)\epsilon^{ij}X_j^N\eta_{MN}\\
&=\frac{1}{2}\int d\tau (\partial_\tau X_i^M-\epsilon_{ik}A^{kl}X_l^M)\epsilon^{ij}X_j^N\eta_{MN}\\
&=\int d\tau \left(\dot{X}^MP_M-\frac{1}{2}A^{ij}X_i^MX_j^N\eta_{MN}\right).
\end{split}
\end{align}
The dot denotes the time derivative with respect to $\tau$. $A_{ij}$ is the $Sp(2, R)$ gauge potential. The equation of motion of this potential goes to the constraints, the target spacetime (1T) Ref.~\cite{survey}\par
We will take the variation of this action to get the equation of motion for the gauge fields and for X, P (Note that:
$
\delta(X^2)=\delta(X^MX_M)=X_M\delta X^M+X^M\delta X_M=2X^M\delta X_M
$, $\delta(X.P)=\delta(X^Mp_M)=P_M\delta X^M+X^M\delta P_M=P^M\delta X_M+X^M\delta P_M$).

{\footnotesize 
\begin{align}
\begin{split}
\delta S&=\int d\tau \begin{pmatrix}
P_M\delta\dot{X}^M+\dot{X}^M\delta P_M-\frac{1}{2}X^2\delta A^{11}-\frac{1}{2}X.P\delta A^{12}-\frac{1}{2}P.X\delta A^{21}-\frac{1}{2}P^2\delta A^{22}\\
-\frac{1}{2}A^{11}\delta (X^2)-\frac{1}{2}A^{12}\delta(X.P)-\frac{1}{2}A^{21}\delta(P.X)-\frac{1}{2}A^{22}\delta (P^2)
\end{pmatrix}\\
&=\int d\tau
\begin{pmatrix}
P_M\delta\dot{X}^M+\dot{X}^M\delta P_M-\frac{1}{2}X^2\delta A^{11}-\frac{1}{2}X.P\delta A^{12}-\frac{1}{2}P.X\delta A^{21}-\frac{1}{2}P^2\delta A^{22}\\
-A^{11}X^M\delta X_M-A^{12}P^M\delta X_M-A^{21}X^M\delta P_M -A^{22}P^M\delta P_M
\end{pmatrix}\\
&=\int d\tau (-\frac{1}{2}X^2\delta A^{11}-\frac{1}{2}(X.P+P.X)\delta A^{12}-\frac{1}{2}P^2\delta A^{22})+\int d\tau \frac{d}{d\tau}(\delta X^MP_M)\\
&+\int d\tau (\dot{X}^M\delta P_M-\dot{P}_M\delta X^M)+\int d\tau (-A^{11}X^M\delta X_M-A^{12}P^M\delta X_M-A^{21}X^M\delta P_M -A^{22}P^M\delta P_M)\\
&=\int d\tau (-\frac{1}{2}X^2\delta A^{11}-\frac{1}{2}(X.P+P.X)\delta A^{12}-\frac{1}{2}P^2\delta A^{22})+\int d\tau \frac{d}{d\tau}(\delta X^MP_M)\\
&+\int d\tau (\dot{X}^M-A^{12}X^M-A^{22}P^M)\delta P_M-\int d\tau (\dot{P}^M+A^{11}X^M+A^{12}P^M)\delta X_M.
\end{split}
\end{align}}
Set $\delta S=0$, we have the equation of motion for the gauge field $A^{11},A^{12},A^{22}$ as
\begin{equation}
X^2=P^2=X.P+P.X=0.
\end{equation}
These are the constraints for Sp(2,R) generators (The generators of Sp(2,R) is in general a function of X and P, $Q_{ij}(X,P)$. In flat spacetime with no background field, they are $Q_{11}=X^2$, $Q_{12}=Q_{21}=X.P+P.X$ and $Q_{22}=P^2$). The above equations have nontrivial solutions only if there are two times, with fewer times these conditions collapse the system to triviality. To see why, consider at first with no time dimension, and ask if there are solutions to equation such as $X^2+Y^2+Z^2+W^2=0$ when there are no minus sign for time. Clearly, this equation just has trivial solution $X^M=P^M=0$ which has no physical content. Next consider the case with one time dimension, we see that the only solution is that $X^M\sim P^M$ must be parallel light-like (null-like) vectors proportional to each other. The angular momentum in defined as 
\begin{equation}
L^{MN}=X^MP^N-X^NP^M,
\end{equation}
will vanish, and hence this is a trivial solution that does not represent even free motion in 1T-physics. In the case of two time dimensions, these equations permit an infinite number of non-trivial solutions. Evidently, the extra time is not introduced "by hand" but the symmetry demands its existence. The number of spatial dimensions is not limited, so in general, the spacetime of 2T-physics is $d+2$ dimensions where d is the number of spatial dimensions.
\\ 
Meanwhile, the equations of motion for X,P are
\begin{equation}
\begin{cases}
\dot{X}^M-A^{12}X^M-A^{22}P^M=0\\
\dot{P}^M+A^{11}X^M+A^{12}P^M=0.
\end{cases}
\end{equation}

The particle model in 2T is still like the standard model \cite{bars2006}, but the space-time is larger than the standard model \cite{bars2006}. Another important thing, the particle model in 2T includes an extra Dilaton in the Higgs potential \cite{bars2006}.

Dilaton is a scalar field which lives in 4+2 dimensions of spacetime. In the standard model of 2T-physics \cite{bars2006}, in order to initiate the electroweak phase transition that explains the source of mass for all massive matter, the 4 + 2 standard model requires the presence of a dilaton field coupled to the Higgs field. The dilaton is required also by the 2T formulation of general relativity in 4 + 2 dimensions. More details about dilaton in 2T-physics can be found in Ref.~\cite{bars2006}.

\subsection{The Higgs-Dilaton potential and gauge fixing technique}
The Higgs-dilaton potential has the following form which is as a result of b-gauge symmetry, the inspiration of this symmetry
comes from BRST formalism \cite{bars2006,kuo2006} but it ultimately comes from the underlying Sp(2,\textbf{R}),
\[
V(\Phi,H) = \la\left(H^\dagger H - \al ^2 \Phi^2\right)^2 + V(\Phi)\, ,
\]
where $\la, \al $ are dimensionless couplings. $H$ and $\Phi$ are the SU(2) Higgs and dilaton doublet respectively. $V(\Phi)$ is the dilaton potential and  plays a role in construction of the effective potential. 

The more complicated work will be done with the Higgs-dilaton  Lagrangian,
\[
L(A,H,\Phi) = \dfrac{1}{2}\Phi\partial^2\Phi + \dfrac{1}{2}\Big(H^\dagger D^2H + (D^2H)^\dagger H \Big) - \la\left(H^\dagger H -
\al ^2 \Phi^2\right)^2 - V(\Phi)\,.
\]
It is convenient if one chooses a lightcone basis in 4+2 dimensions written as
\begin{align}
&X^{\pm '}_i = \dfrac{1}{\sqrt{2}}\left(X^{0'}_i\pm X^{1'}_i\right), \text{where } X_i^M = (X^M_1,X^M_2) \equiv (X^M, P^M)\\
\Rightarrow&
\begin{cases}
X^2 = - 2X^{+'}X^{-'} + X^\mu X_\mu ,\\
X^MP_M = -X^{+'}P^{-'} - X^{-'}P^{+'} + X^\mu P_\mu\, .
\end{cases}
\end{align}

We choose the Bars's parametrization
\cite{bars2000a} for the components of $X^M$ as follows
\begin{align}
X^{+'} = \ka , X^{-'} = \ka \La, X^\mu = \ka  x^\mu \label{3.1}\\
\Rightarrow \ka  = X^{+'}, \La = \dfrac{X^{-'}}{X^{+'}}, x^\mu = \dfrac{X^\mu }{X^{+'}}\, ,
\end{align}
from this choose (the gauge fixing technology), we can reduce the 2T metric to 1T metric (Minkowski metric). 

Scalar fields will be reduced to the following:
\begin{align}\label{3.2a}
\begin{cases}
\Phi(X) \longrightarrow \dfrac{1}{\ka }\phi(x)\quad ;\quad H(X) \longrightarrow \dfrac{1}{\ka }h(x) , \\
\partial^2 \Phi(X) = \partial^M\partial_M \Phi(X) \longrightarrow \dfrac{1}{\ka ^3}\dfrac{\partial^2 \phi(x)}{\partial x^\mu \partial x_\mu}\, ,\\
D^2 H(X) = D^M D_M H(X) \longrightarrow \dfrac{1}{\ka ^3}D^\mu D_\mu h(x)\, .
\end{cases}
\end{align}

With the above gauge fixing, one would be able to derive the following reduction
\begin{align}
L(A,H,\Phi) \longrightarrow \ &\dfrac{1}{2\ka ^4}\phi\dfrac{\partial^2\phi}{\partial x^\mu\partial x_\mu}
+ \dfrac{1}{2\ka ^4}\left[h^\dagger D_\mu  D^\mu h + (D^\mu D_\mu h)^\dagger h\right]\crn
& \qquad\qquad\qquad\qquad\qquad\qquad- \dfrac{\la}{\ka ^4}\left(h^\dagger h - \al ^2\phi^2\right)^2 - V(\phi)\,.
\end{align}

Note that all the $\kappa$ coefficients will be simplified by the scaling invariant of the actions through $\delta(X^2)$ when reducing from 2T to 1T \cite{bars2006,kuo2006}. The Higgs and Dilaton have redefined \cite{bars2006,kuo2006} as

\begin{equation}
h(x) = \dfrac{1}{\sqrt{2}}\begin{pmatrix}
0\\
v+\si(x)
\end{pmatrix}; \langle h(x)\rangle = v\,; \phi(x)=\dfrac{1}{\al \sqrt{2}}(v + \al  d(x));\langle\phi(x)\rangle = \dfrac{v}{\al}\,.
\end{equation}

As discussed in Refs.~\cite{bars2006,kuo2006}, there are some reasons to expect that the new dilaton potential only takes the form $V(\Phi)=\rho\Phi^4$ and after gauge fixing to 1T-physics it takes the form $V(\phi)=\rho\phi^4$. We assume that the resulting scalar field after reducing the dilaton field from 2T-physics to 1T-physics is indeed the inflaton field, and we will use this quartic form of the potential
	\begin{equation}\label{49}
	V(\phi)=\rho \phi^4
	\end{equation}
to study warm inflation in the following sections and notice that $\rho$ has a very small value, which will be explained in a later section.

The minimum process of this full Higgs-Dilaton potential can be considered as the minimum process of the potential when not $V(\phi)$. Because $\rho$ is very small. The minimum process is detailed in Ref.\cite{bars2006}.

The full Higgs-Dilaton potential is expanded as follows

\begin{align}
V(h,\phi) &= \dfrac{\lambda}{4}\left[(v + h)^2 - (v + \alpha d)^2\right]^2 + \dfrac{\rho}{4\alpha^4}(v + \alpha d)^4 \nonumber\\
&= \dfrac{\lambda}{4}(h-\alpha d)^2 (2v + h +\alpha d)^2\nonumber \nonumber\\
&\qquad+\dfrac{\rho}{4\alpha^4}(v^4 + 4\alpha v^3 d + 6\alpha^2v^2 d^2 +4\alpha^3 v d^3 + \alpha^4d^4) \nonumber\\
&= \dfrac{\lambda}{4}(4v^2 h^2 + 4\alpha^2v^2 d^2- 8\alpha v^2 h d)\nonumber \nonumber\\
&\qquad + \dfrac{\rho}{4\alpha^4}(v^4 + 4\alpha v^3 d+ 6\alpha^2v^2 d^2) + \text{3,4-fields interaction terms}\\
&= \dfrac{\lambda}{4}(4v^2 h^2 + 4\alpha^2v^2 d^2- 8\alpha v^2 h d)+ \frac{3\rho}{2\alpha^2}v^2 d^2\nonumber \nonumber\\
&\qquad + \dfrac{\rho}{4\alpha^4}(v^4 + 4\alpha v^3 d) + \text{3,4-fields interaction terms}\\
&=v^2(h\quad d).\begin{pmatrix}
	 \lambda& -\alpha\lambda\\
	-\alpha\lambda & \lambda\alpha^2+\frac{3\rho}{2\alpha^2}
\end{pmatrix}.\begin{pmatrix}
h\\
d
\end{pmatrix}+ \dfrac{\rho}{4\alpha^4}(v^4 + 4\alpha v^3 d) + \text{3,4-fields terms}\\
&\approx\frac{1}{2}(h'\quad d').\begin{pmatrix}
	m^2_{h'}& 0\\
	0 & m^2_{d'}
\end{pmatrix}.\begin{pmatrix}
	h'\\
	d'
\end{pmatrix}+ const + \text{3,4-fields terms},\label{21}
\end{align}

in which
\bea
m^2_{d'}=\frac{v^2 \left(2 \alpha ^4 \lambda +2 \alpha ^2 \lambda -\sqrt{\left(2 \alpha ^4 \lambda +2 \alpha ^2 \lambda +3 \rho \right)^2-24 \alpha ^2 \lambda  \rho }+3 \rho \right)}{\alpha ^2},\label{21a}
\eea
\bea
m^2_{h'}=\frac{v^2 \left(2 \alpha ^4 \lambda +2 \alpha ^2 \lambda +\sqrt{\left(2 \alpha ^4 \lambda +2 \alpha ^2 \lambda +3 \rho \right)^2-24 \alpha ^2 \lambda  \rho }+3 \rho \right)}{\alpha ^2}.\label{21b}
\eea

The physical fields are
\bea
\begin{pmatrix}
	h'\\
	d'
\end{pmatrix}=\left(
\begin{array}{cc}
	A & B\\
	1 & 1 \\
\end{array}
\right).\begin{pmatrix}
	h\\
	d
\end{pmatrix},
\eea
\bea
A=\frac{2 \alpha ^4 \lambda -2 \alpha ^2 \lambda +\sqrt{\left(2 \alpha ^4 \lambda +2 \alpha ^2 \lambda +3 \rho \right)^2-24 \alpha ^2 \lambda  \rho }+3 \rho }{4 \alpha ^3 \lambda },
\eea
\bea
B=-\frac{-2 \alpha ^4 \lambda +2 \alpha ^2 \lambda +\sqrt{\left(2 \alpha ^4 \lambda +2 \alpha ^2 \lambda +3 \rho \right)^2-24 \alpha ^2 \lambda  \rho }-3 \rho }{4 \alpha ^3 \lambda }. 
\eea

To do the above minima and approximation, we have to remove the term $\dfrac{\rho v^3 d}{\alpha^3}$ or $\dfrac{\rho v^3}{\alpha^3}\ll 1$. In addition, the angle of mixing between $h$ and $\phi$ is proportional to $\alpha$, which is significant in an inflationary scenarino. With large $\alpha$, $\phi$ couples to the standard model particles as large as $h$ does. The signal strength that produces $\phi$ at the LHC, for example by gluon fusion, is comparable to that of the Higgs field \cite{them1, them2}. Therefore, the large $\alpha$ situation must be omitted.

From Eqs.~\ref{21a} and \ref{21b}, when $\rho$ goes to zero, we derive the massless Dilaton, this is consistent with the discussion in Ref.~\cite{bars2006,kuo2006}.
 
\section{Dilaton as inflaton in the warm $\phi^4$ inflation}\label{sec3}

Inflation is a very short period of exponential expansion in the early Universe. This era is assumed to happen in order to explain some problems of the standard Big Bang cosmology including the horizon problem, the flatness problem, and the seeds of cosmic structures. The observational evidences strongly support the general ideas of inflation, but the details of inflation remain elusive.

In standard inflation, there are two evident ways to calculate the  inflaton mass. The first one is using the quadratic inflaton potential
$V=\fr{1}{2}m^2\varphi^2$, where $m$ is the  inflaton mass. The second one is using the exponential potential $V=e^{-\beta\varphi}$.
In the latter case, we expand the Taylor series of the potential and the term in front of $\varphi^2$ is the inflaton mass.

Warm inflation is different to the standard inflation, is a model of inflation in which, apart from the quantum fluctuations of the inflaton field, the inflaton field is also subject to thermal fluctuations due to the interaction with the background thermal bath. This idea was first initiated by Berrera in 1995 \cite{Ber95}. Intuitively, we expect that in this model the inflationary gravitational waves amplitudes must be smaller than that in the standard "cold" inflation. The reason is the following. The kinetic energy of the inflaton field will be dissipated due to the additional thermal fluctuations and hence its roll is slowed down further. This means that the inflaton field excursion can be small and yet still produces enough the number of e-folds to resolve the horizon or flatness problems. In turn, this implies that the tensor-to-scalar ratio is also small due to the Lyth bound. Indeed, we will see that this expectation turns out to be true. The $\phi^4$ potential in the standard "cold" inflation is ruled out since it predicts a too large tensor-to-scalar ratio; on the other hand, the $\phi^4$ potential in warm inflation works perfectly well since the predicted tensor-to-scalar ratio in this model is much smaller.

However, both of them use the slow rolling approximation (Slow-roll approximation states that $\fr{1}{2}\dot{\varphi}^2 \ll V$, and therefore $\ep ,|\eta| \ll 1$. Slow-roll approximation is the must for inflation to happen. There are two	formalisms of slow-roll approximation which are potential slow-roll approximation (PSRA) and Hubble slow-roll approximation (HSRA). In this paper, we will always use the former). The slow-roll parameters are defined as \cite{39}
\[
\ep =\fr{M_p^2}{2}\left(\fr{V'}{V}\right)^2\hs;\hs \eta=M_p^2\fr{V''}{V}\,,
\]
 where $M_p$ is 4D Planck scale and primes denote derivatives with respect to $\va$.  

In warm inflation, we have two cases which are the weak and the strong dissipative regimes  defined as $R\ll 1$ and $R\gg 1 $, respectively:
\begin{equation}
R=\frac{\Gamma}{3H},
\end{equation}
where $H$ is the Hubble parameter and $\Gamma$ is the inflaton decay rate. For the warm inflation, in addition to the two usual slow-roll parameters, we have the following two parameters \cite{wr1,wr2,wr3,phi4a,phi4b,phi4c},
\begin{equation}
\beta=M_p^2\fr{\Gamma' V'}{\Gamma V}, \sigma=M_p^2\fr{V'}{\varphi V}.
\end{equation}
The slow-roll conditions can be expressed as \cite{wr1,wr2,wr3,phi4a,phi4b,phi4c}
\begin{equation}
\epsilon, |\eta|, |\beta|,|\sigma| \ll 1+R.
\end{equation}

 In order to make sure that the potential $V\sim\va^4$ is still reliable, we need to calculate two important quantities, which are the scalar spectral index $n_s$ and the tensor-to-scalar ratio $r$. To do that, we first need to mention an important quantity in inflation which is the e-folding number $N$ representing the amount of inflation. In the context of potential slow-roll approximation, the e-folding number is
\be\label{e-folding1} 
N\equiv ln\fr{a(t_f)}{a(t_i)}\simeq -\fr{8\pi}{M_4^2}\int_{\va_i}^{\va_f}\fr{V(1+R)}{V'}d\va\, .
\ee
From the condition of inflation to end $\ep \sim 1+R$, we get the final value of the inflaton field $\va_f$. Then, using Eq.~(\ref{e-folding1})
to calculate the initial value of the inflaton field $\va_i$ when the perturbations exit the horizon. Therefore, $n_s$ and $r$ can be expressed in terms of the slow-roll parameters and assuming an inflaton decay rate $\Gamma=aT$.  Finally, with the inflation at 60 e-folds, one gets $10^{-15}<\rho<10^{-13}$ in all case \cite{wr1,wr2,wr3,phi4a,phi4b,phi4c}.

Considering our assumption in \eqref{49} to fit the $SP(2,R)$ symmetry, the Dilaton potential has the quartic form. Thus, this form is compatible with a warm inflation scenario. A quartic inflaton potential in the warm inflation scenario has also been studied in \cite{wr1,wr2,wr3,phi4a,phi4b,phi4c}. 

Furthermore, the determination of the value of dissipative coefficient $\Gamma=aT$ (or $a$) currently only stops at the estimate in the example models. Because the determination of the dissipative components of Inflaton is not specific. This can go beyond the standard model and SM-like versions as 2T model. Calculations in some specific cases and the expansion of $\Gamma$ can be found in Refs.~\cite{wr1, 22, 23, 24, 25, 27, 28}. However, we can roughly outline that $a$ is a function of coupling constants of the dissipative channels of Inflaton.

Based on Ref.~\cite{phi4b}, with strong and weak dissipative regime, the important quantities are summarized in Table \ref{2}. Especially the values of $\rho$ (or $\lambda$) and $a $ is derived when combined with Planck data \cite{30planck}.

\begin{table}[htp]
	\centering 
	\begin{tabular}{ c||c|c } 
		\hline
		\hline
		\diagbox{Quantity}{Regime} & Weak dissipative regime & Strong dissipative regime \\ 
		\hline
		\hline
		The power spectrum:$P_R^{1/2}\sim some\times 10^{-5}$ & $\left(\frac{\lambda' \sqrt{a} N^3}{6\sqrt{70}\pi^3}\right)^{1/3}$ &$\left[\frac{4 N^3 \lambda'}{125 \pi^{8/3}}\left(\frac{2}{315}\right)^{1/3}\right]^{3/8}$ \\ 
		
		The tensor-to-scalar ratio: $r(n_s)$ & $\frac{4\sqrt{14}}{625 \sqrt{5}\,a^{1/2}}(1-n_s)$ & $8.5\times 10^{-9}\frac{\,\pi^{10/3}}{a^4}(1-n_s)$ \\
		
		$\rho=\frac{\lambda'}{4}$ & $0.25\times 10^{-15}<\rho<0.25\times 10^{-13}$ &$some\times 10^{-15}$ \\
		
		$a$ & $ 6.5\times 10^{-5}< a<3.4\times 10^{-2}$&$3.4\times 10^{-2}< a$\\		
		\hline
		\hline
	\end{tabular}
	\caption{The values of quantities are calculated to give the result $r(n_s)$ that satisfy the Planck data \cite{30planck}. $n_s$ is the scalar spectral index.}
	\label{2}
\end{table}

In the Table \ref{2}, to satisfy all cases, we choose $\rho=10^{-14}$.
Thus, in the weak dissipative regime, $a=5\times 10^{-4}$. In the strong dissipative regime, we choose $a=5\times 10^{-2}$. 

With $m_{h'}=125$ GeV, $v=246$ GeV and $\rho=10^{-14}$, we can deduce the following relationship
\bea
\la=\frac{3.05176\times 10^{21} \alpha ^2-7.09172\times 10^8}{4.72781\times 10^{22} \alpha ^4+4.72781\times 10^{22} \alpha ^2-1.09866\times 10^{10}}.
\eea

Furthermore, we have the condition,
\bea
\frac{\rho v^3}{\alpha^3}=\frac{10^{-14}\times 246^3}{\alpha^3}\ll 1 \Longrightarrow \alpha \gg 0.0053.
\eea

\begin{figure}[htbp]
	\centering
	\includegraphics[width = 0.6\textwidth]{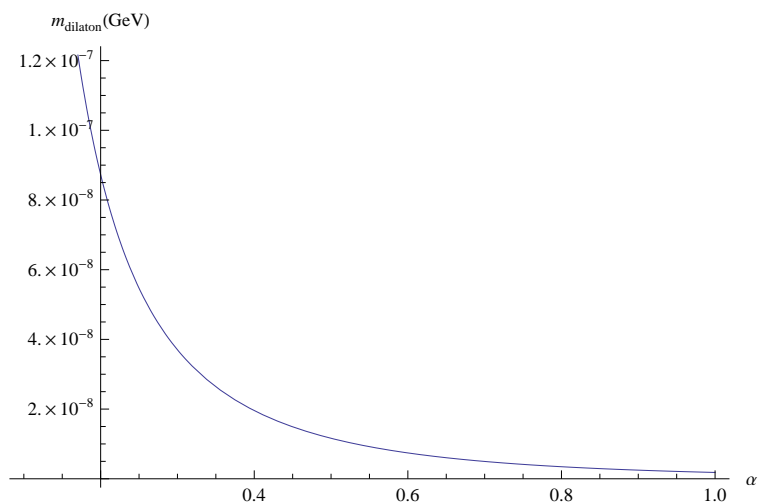}
		\caption{The mass of Dilaton with $\rho=10^{-14}$.}
	\label{fig:3.13}
\end{figure}

\begin{figure}[htbp]
	\centering
	\includegraphics[width = 0.6\textwidth]{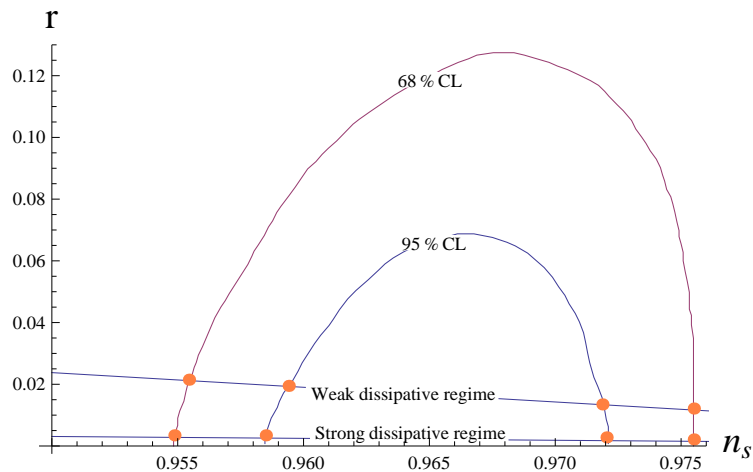}
	\caption{The function $r$ depends on $n_s$. $\rho=10^{-14}$, $0.1<\alpha<1$.}
	\label{fig:3.14}
\end{figure}

We then deduce the mass of dilaton in Fig.~\ref{fig:3.13}. Because $\alpha$ is small, the mass of Dilaton decreases, must be smaller than $10^{-7}$ GeV when $0.1<\alpha<1$. So Fig.~\ref{fig:3.14} is plotted, to show how $r(n_s)$ is compatible with the Planck data \cite{30planck}.

We see that, the larger the $\alpha$, the smaller the mass of the dilaton. Therefore, the signal of Dilaton will be very small even though the coupling constant with Higgs is large.

Finally, when combined with the Planck data \cite{30planck}, in all regimes we get $\rho=10^{-14}$, from which the coupling constant ($\alpha$) between Higgs and Dilaton can be deduced.

\section{Conclusion and discussion}\label{sec4}

We showed that Dilaton can be a candidate of Inflaton in the warm inflation scenario. Accordingly, the parameter domain of the Higgs-Dilaton potential is retrieved. This indirectly shows that extra dimensions can play an important role in the inflation problem.

The Higgs-Dilaton potential can be added $\phi^2$ that is the dominant component of the standard inflation, thus a dimensional mass parameter can be added along with $\phi^2$. But it breaks the basic $Sp(2;R)$ symmetry. This parameter can be derived from the interaction of Dilaton with $S(x)$ which is an additional scalar field, can be seen as Inflaton \cite{1bars, 2bars, 2barsb}. This has been done consistently with the scale invariance required by 2T-Physics Ref.~\cite{3bars}.

According to Refs.~\cite{wr1, 22, 23, 24, 25, 27, 28}, to be able to find more dissipative components in 2T physics, we need to extend this model in combination with supersymmetry or extending the Higgs-Dilaton potential one more time with exotic particles but must conform to the $Sp(2,R)$ symmetry. Also, because in the 2T model, Dilaton only interacts directly with Higgs. Therefore, there are not enough dissipative components to have a suitable value of $a$ or $r$ for the Planck data \cite{30planck}. This is an interesting work after this article.

There are the other studies in Cosmology, or example, the cosmological constant and inflation \cite{42a,42b}. But in this model, the inflation and reheating period have not been analyzed in detail. We considered that Dilaton is Inflaton in the warm scenario. This presents another opportunity to study extra dimensions as well as other inflationary scenarios.

The extra dimensional phenomena can give many prediction for experiments. Our next work focuses on extra dimensions, in order to describe their properties in the baryogenesis or leptongenesis scenario and cosmology problems.

The 2T model is a hologram of string theory and has a connection with SM. Therefore, analyzing and refining this model is a "road" to create a good bridge between two theories, in which the problem of inflation and the matter-antimatter asymmetry are significant problems.

New physical effects from the interaction between Higgs and Dilaton may change the electroweak, but this model still fits into the standard model through the Bars transformation that has been mentioned in section II.2. Thus, these changes (such as the Higgs couplings to fermions)  will indeed provide new explanations for difficult problems (such as Baryogenesis, etc. ) without changing the electroweak nature.

In a context with too many free parameters, the first step in our paper shows that the small Dilaton potential can be combined with the warm inflation problem, and then derive the parameter domain $\alpha$. If further extended to other problems, the range of values $\alpha$ that is derived in this paper, need to be confirmed more fully.

\section*{ACKNOWLEDGMENTS}

In memory of the day our venerable teacher, Dr. Vo Thanh Van, passed away on 5 July 2016, this series of our papers highlights the continuation of what he has done through many years in the Department of Theoretical Physics.

\end{document}